\documentclass[twocolumn,prd,superscriptaddress,preprintnumbers,nofootinbib]{revtex4}[11pt]
\pdfoutput=1
\usepackage{amsmath,amssymb,graphicx}
\graphicspath{{figs/}}
\usepackage{epsf,verbatim}
\usepackage{hyperref}
\usepackage{comment}
\usepackage{color}
\usepackage{slashed}
\usepackage{subfigure}
\usepackage[usenames,dvipsnames]{xcolor}
\usepackage{comment}
\usepackage{mdwlist, paralist}
\usepackage{rotating}
\usepackage{multirow}

\def\to{\rightarrow}
\def\ord{\mathcal{O}}
\def\TeV{~{\mbox{TeV}}}

\def\GeV{~{\mbox{GeV}}}

\def\mL{\mathcal{L}}

\def\bM{\mathbb{M}}

\begin{document}

\title{\boldmath On the $ZH\eta$ vertex in the simplest Little Higgs Model}


\affiliation{Center for Future High Energy Physics \& Theoretical Physics Division,
Institute of High Energy Physics, Chinese Academy of Sciences, Beijing 100049, China}
\affiliation{Physics Division, National Center for Theoretical Sciences, Hsinchu, Taiwan 300}
\affiliation{Institute of Theoretical Physics \& State Key Laboratory of Nuclear Physics and Technology, Peking University, Beijing 100871, China}
\affiliation{Collaborative Innovation Center of Quantum Matter, Beijing 100871, China}
\affiliation{Center for High Energy Physics, Peking University, Beijing 100871, China}

\author{Shi-Ping He}
\thanks{sphe@pku.edu.cn}
\affiliation{Institute of Theoretical Physics \& State Key Laboratory of Nuclear Physics and Technology, Peking University, Beijing 100871, China}

\author{Ying-nan Mao}
\thanks{maoyn@ihep.ac.cn}
\affiliation{Center for Future High Energy Physics \& Theoretical Physics Division,
Institute of High Energy Physics, Chinese Academy of Sciences, Beijing 100049, China}

\author{Chen Zhang}
\thanks{czhang@cts.nthu.edu.tw}
\affiliation{Physics Division, National Center for Theoretical Sciences, Hsinchu, Taiwan 300}
\affiliation{Institute of Theoretical Physics \& State Key Laboratory of Nuclear Physics and Technology, Peking University, Beijing 100871, China}

\author{Shou-hua Zhu}
\thanks{shzhu@pku.edu.cn}
\affiliation{Institute of Theoretical Physics \& State Key Laboratory of Nuclear Physics and Technology, Peking University, Beijing 100871, China}
\affiliation{Collaborative Innovation Center of Quantum Matter, Beijing 100871, China}
\affiliation{Center for High Energy Physics, Peking University, Beijing 100871, China}

\begin{abstract}
The issue of deriving $ZH\eta$ vertex in the simplest Little Higgs (SLH) model is revisited. Special attention
is paid to the treatment of non-canonically-normalized scalar kinetic matrix and vector-scalar two-point
transitions. We elucidate a general procedure to diagonalize a general vector-scalar system in gauge theories
and apply it to the case of SLH. The resultant $ZH\eta$ vertex is found to be different from those which have
already existed in the literature for a long time. We also present an understanding of this issue from
an effective field theory viewpoint.
\end{abstract}

\maketitle

\setcounter{equation}{0} \setcounter{footnote}{0}

\section{Introduction}

The discovery of the $125\GeV$ Higgs-like boson~\cite{Aad:2012tfa,Chatrchyan:2012xdj}
marks a prominent triumph of the Standard Model (SM). Nevertheless, it is widely
believed that this is not the end of the story. The SM in its current form leaves
too many unanswered questions, from theoretical ones like the issue of Higgs mass
naturalness~\cite{Giudice:2008bi,Giudice:2013nak}, to observational ones like the
nature of the dark matter present in the
universe~\cite{Bertone:2004pz,Lisanti:2016jxe}. Almost all models going beyond
the SM (BSM) entail an enlargement of the scalar sector, and consequently forms
of interaction which are absent in the SM could be possible. Searching for such
kind of new interactions therefore may lead to decisive evidence of the
existence of BSM and provide a clue to the nature of the BSM physics.

For example, Lorentz symmetry does not forbid the interaction of one gauge boson (denoted as $Z$) with two
scalar bosons (denoted as $H$ and $\eta$) at the dimension-4 level, in the form like
\begin{equation}
Z^\mu(H\partial_\mu\eta-\eta\partial_\mu H)
\label{eq:aszheta}
\end{equation}
The SM has only one Higgs particle and thus cannot accommodate such kind of vector-scalar-scalar (VSS) interactions\footnote{Here we mean physical fields. Unphysical fields like Goldstone or ghost can certainly participate in VSS interactions in the SM.}. Going beyond
the SM, the appearance of interactions like Eq.~\eqref{eq:aszheta} is quite common in models like the
two-Higgs-doublet model (2HDM) and supersymmetric models, which may lead to the associated production of two
scalar bosons~\cite{Kanemura:2001hz,Cao:2003tr} or Higgs-to-Higgs cascade decays~\cite{Coleppa:2014hxa,Coleppa:2014cca} as important collider signatures.

Besides the usual 2HDM and supersymmetric models which contain a linearly-realized scalar sector, VSS interactions
have also been studied in the context of nonlinearly-realized scalar sectors. Nonlinearly-realized scalar sectors
are frequently adopted when building a model in which the Higgs is realized as a pseudo-Goldstone boson of some
global symmetry breaking~\cite{Bellazzini:2014yua}, which could be helpful in addressing the hierarchy problem. In principle the derivation
of VSS vertices in such models is similar to the linearly-realized case: start from the gauge covariant kinetic
terms of the scalar fields and then expand the interaction fields into vacuum expectation values and mass
eigenstate fields after which the three-point VSS vertices could be extracted. Nevertheless there can be important
technical differences in intermediate steps. When the scalar sector is nonlinearly-realized, scalar kinetic terms are in general not
automatically canonically normalized, and there can be ``unexpected" vector-scalar two-point transitions which
need to be taken care of. We will show in the following sections that these situations indeed occur for the
case of the simplest Little Higgs (SLH) model~\cite{Schmaltz:2004de}, which is proposed as a simple solution to the Higgs mass
naturalness problem.

From a more general perspective, the problem we encounter is how to diagonalize a vector-scalar system
in gauge field theories. Specifically, the Lagrangian we start with might not be canonically normalized
in its kinetic part, and may have some general vector-scalar two-point transitions. To do perturbation
theory in the usual manner, we need to first render its kinetic part canonically normalized, which could
be done via the usual complete-the-square method. To remove the vector-scalar two-point transitions,
strictly speaking we need to choose appropriate gauge-fixing terms. Finally we still need to diagonalize
the scalar mass matrix with contribution from both the original scalar mass terms and the gauge-fixing
terms. These steps set the stage for the derivation of VSS interactions.

In Section~\ref{sec:sys} the systematic procedure of diagonalize a general vector-scalar system in
gauge field theories will be elucidated. Then in Section~\ref{sec:slh} we apply this procedure to
the SLH model and derive the mass eigenstate $ZH\eta$ vertex\footnote{By `mass eigenstate' $ZH\eta$
vertex we mean the $ZH\eta$ vertex obtained after rotating $Z,H,\eta$ fields into their corresponding
mass eigenstates. For previous studies related to the $\eta$ particle in the SLH,
we refer the reader to ~\cite{Kilian:2004pp,Kilian:2006eh,Han:2005ru,Cheung:2006nk,Cheung:2007sva,
Cheung:2008zu,Lu:2007jj,Han:2009zp,Wang:2010fh,Wang:2011qz,Kim:2011bv,Han:2013ic,Mao:2017hpp}.}
to $\ord\left((\frac{v}{f})^3\right)$. The $ZH\eta$ vertex
derived here is found to be different from those which have already existed in the
literature~\cite{Kilian:2004pp,Kilian:2006eh} for
a long time. In Section~\ref{sec:conc} we present our discussion and conclusion.

\section{General diagonalization procedure}
\label{sec:sys}

Consider a gauge field theory in which there are $n_S$ real scalar fields $G_i,i=1,2,...,n_S$
and $n_M$ real massive gauge boson fields $Z_p^{\mu},p=1,2,...,n_M$. If complex fields exist,
we can always decompose them into their real components and proceed in a similar manner. The
$G_i$'s which we start with neither need to be canonically normalized nor need to have
diagonalized mass terms. For simplicity (but without loss of generality) the $Z_p$'s are assumed
to have canonically normalized kinetic terms but don't have to be diagonalized in their mass terms.
When we say the $Z_p$'s are massive, it means that the eigenvalues of the mass matrix of
$Z_p$'s are all positive. Especially, massless gauge bosons like photon are temporarily
excluded from discussion. However, generalizing the procedure to theories containing
massless gauge bosons is straightforward.

Now suppose the classical Lagrangian of this gauge theory contains the following quadratic
parts\footnote{Here we suppress the gauge boson kinetic terms which
are assumed to be already canonically normalized.}
(summation over repeated indices is implicitly assumed):
\begin{widetext}
\begin{equation}
\mL_{quad}\supset\frac{1}{2}V_{ij}(\partial_\mu G_i)(\partial^\mu G_j)
+F_{pi}Z_p^\mu(\partial_\mu G_i)-\frac{1}{2}(\bM_G^2)_{ij}G_i G_j
+\frac{1}{2}(\bM_V^2)_{pq}Z_{p\mu}Z_q^\mu
\label{eq:quad}
\end{equation}
\end{widetext}
Here $V$ is a real invertible $n_S\times n_S$ symmetric matrix, $F$
is a real $n_M\times n_S$ matrix, $\bM_G^2$ is a $n_S\times n_S$
symmetric matrix the rank of which does not exceed $n_E\equiv n_S-n_M$
\footnote{Here we assume all the $Z_p$'s acquire their masses by eating
appropriate Goldstones. In compliance with the fact that $n_M$ massless
Goldstones should exist before gauge-fixing, the rank of $\bM_G^2$
should not exceed $n_S-n_M$.}, and
$\bM_V^2$ is a real $n_M\times n_M$ symmetric matrix which has $n_M$ positive
eigenvalues. The elements of the four matrices $V,F,\bM_G^2,\bM_V^2$ depend
only on the model parameters, not on field variables. For convenience let us
define
\begin{equation}
\tilde{G}_p=F_{pi}G_i,\quad p=1,2,...,n_M
\end{equation}
Then the vector-scalar two-point transition term (the second term on
the right hand side of Eq.~\eqref{eq:quad}) is simply
$Z_p^\mu\partial_\mu\tilde{G}_p$.

To carry out perturbation theory, it is preferable to eliminate the
vector-scalar two-point transitions, make the scalar kinetic terms
canonically normalized and at the same time diagonalize the scalar
and vector mass terms. We will see that the procedure involved
actually goes hand in hand with the quantization of the theory.
Also, the tight structure of the gauge theory greatly facilitates
the diagonalization process.

In gauge field theories, the vector-scalar two-point transitions
are usually eliminated by adding appropriate gauge-fixing terms.
If we require the $R_\xi$ gauge-fixing procedure remove
all the vector-scalar two-point transitions, then it is natural
to consider adding the following gauge-fixing Lagrangian:
\begin{equation}
\mL_{gf}=-\sum_{p=1}^{n_M}\frac{1}{2\xi^p}(\partial_\mu Z_p^\mu-\xi^p\tilde{G}_p)^2
\label{eq:gf}
\end{equation}
Here $\xi^p,p=1,2,...,n_M$ are gauge parameters. There is freedom in the choice
of the gauge-fixing function and the requirement to remove vector-scalar two-point
transitions is not sufficient to uniquely determine it.
However we will see below there is a theoretically well-motivated choice
which facilitates the diagonalization process. After adding the gauge-fixing terms,
we have
\begin{align}
& \mL_{quad}+\mL_{gf}\supset
\frac{1}{2}V_{ij}(\partial_\mu G_i)(\partial^\mu G_j)
-\frac{1}{2}\xi^p\tilde{G}_p^2\nonumber \\
&-\frac{1}{2}(\bM_G^2)_{ij}G_i G_j
-\frac{1}{2\xi^p}(\partial_\mu Z_p^\mu)^2+\frac{1}{2}(\bM_V^2)_{pq}Z_{p\mu}Z_q^\mu
\end{align}

The matrix $V$ denotes the scalar kinetic matrix. If it is not the identity
matrix, we may simply use the complete-the-square method to diagonalize
it and then make the resulting terms canonically normalized. This is in
complete analogy to the diagonalization of quadratic forms in linear algebra.
Note that the overall transformation employed to render the scalar kinetic
terms canonically normalized need not be orthogonal.

Now suppose we have found a transformation of the scalar fields
\begin{equation}
S_i=U_{ij}G_j
\label{eq:sug}
\end{equation}
which renders the scalar kinetic terms diagonalized and canonically normalized:
\begin{equation}
\frac{1}{2}V_{ij}(\partial_\mu G_i)(\partial^\mu G_j)
=\frac{1}{2}(\partial_\mu S_i)(\partial^\mu S_i)
\end{equation}
Here $U$ is a real invertible $n_S\times n_S$ matrix which only needs to
satisfy
\begin{equation}
V=U^T U
\end{equation}
It is evident that $U$ is not uniquely determined. It is only determined
up to an orthogonal transformation. We may take advantage of this freedom
to do additional orthogonal transformation to further diagonalize the scalar
mass matrix while still keeping scalar kinetic terms in their canonically
normalized form.

After the transformation Eq.~\eqref{eq:sug} we obtain
\begin{widetext}
\begin{align}
\mL_{quad}+\mL_{gf}\supset\frac{1}{2}(\partial_\mu S_i)(\partial^\mu S_i)
-\frac{1}{2}\xi^p\tilde{G}_p^2-\frac{1}{2}((U^{-1})^T\bM_G^2 U^{-1})_{ij}S_i S_j 
-\frac{1}{2\xi^p}(\partial_\mu Z_p^\mu)^2+\frac{1}{2}(\bM_V^2)_{pq}Z_{p\mu}Z_q^\mu
\label{eq:quad1}
\end{align}
\end{widetext}
In the above equation $\tilde{G}_p$'s can be viewed as linear combinations of $S_i$'s.
It should be noted from a physical perspective that the $n_S$ scalar degrees
of freedom with which we started could be divided into two categories (after
appropriate linear combinations if needed): unphysical scalars and physical
scalars. Specifically, $n_M$ unphysical scalars should exist and serve as
unphysical Goldstones to be eaten by $n_M$ gauge bosons to make them massive.
The remaining $n_E=n_S-n_M$ scalar degrees of freedom then must be physical scalars.
By virtue of this observation, there must exist an orthogonal transformation
\begin{equation}
\bar{S}_i=P_{ij}S_j
\label{eq:pt}
\end{equation}
which diagonalizes the $-\frac{1}{2}((U^{-1})^T\bM_G^2 U^{-1})_{ij}S_i S_j$ term.
Then Eq.~\eqref{eq:quad1} becomes
\begin{widetext}
\begin{align}
\mL_{quad}+\mL_{gf}\supset\frac{1}{2}(\partial_\mu\bar{S}_i)(\partial^\mu\bar{S}_i)
-\frac{1}{2}\xi^p\tilde{G}_p^2-\frac{1}{2}\nu_r^2\bar{S}_r^2
-\frac{1}{2\xi^p}(\partial_\mu Z_p^\mu)^2+\frac{1}{2}(\bM_V^2)_{pq}Z_{p\mu}Z_q^\mu
\label{eq:quad2}
\end{align}
\end{widetext}
The index $r$ ranges from $n_M+1$ to $n_S$ (this will be assumed
whenever we use the index $r$), and $\nu_r$'s depend only on model parameters,
not on field variables. With this labeling convention the latter $n_E$ fields in $\bar{S}_i$'s
correspond to physical scalars while the remaining ones are unphysical Goldstone bosons. The
matrix $P$ and the $\nu_r$'s can be made independent of the $\xi^p$'s, because in the course of
diagonalizing the $-\frac{1}{2}((U^{-1})^T\bM_G^2 U^{-1})_{ij}S_i S_j$ term, the
$-\frac{1}{2}\xi^p\tilde{G}_p^2$ term is left untouched.

It is helpful to recall that in Eq.~\eqref{eq:quad2} the $\tilde{G}_p$'s can
be viewed as linear combinations of $\bar{S}_i$'s. In fact, because $n_E$
physical scalars must exist, the matrix $P$
can be chosen so that the $\tilde{G}_p$'s do not contain the $\bar{S}_r$'s.
That is to say, the $\tilde{G}_p$'s can be expressed as linear
combinations of $\bar{S}_i,i=1,2,...,n_M$. Therefore, by examining
Eq.~\eqref{eq:quad2} it is obvious that in $\mL_{quad}+\mL_{gf}$
the $n_E$ physical scalars are clearly separated from the unphysical ones
after the orthogonal transformation Eq.~\eqref{eq:pt}.

At this stage we need to take a closer look at the unphysical scalar mass
term in Eq.~\eqref{eq:quad2}, which is
\begin{equation}
\mL'\equiv-\frac{1}{2}\xi^p\tilde{G}_p^2
\label{eq:lp}
\end{equation}
Recalling that the $\tilde{G}_p$'s are linear combinations of
$\bar{S}_i,i=1,2,...,n_M$, the next thing we need to do is to find an
orthogonal transformation
\begin{equation}
\tilde{S}_i=K_{ij}\bar{S}_j
\label{eq:kt}
\end{equation}
which diagonalizes $\mL'$. In Eq.~\eqref{eq:kt} $i,j$ range from
$1$ to $n_S$, and $K$ is a $n_S\times n_S$ orthogonal matrix.
Nevertheless, to avoid spoiling the already
diagonalized physical scalar mass term, it is advisable
to consider the following block-diagonal form of $K$:
\begin{align}
K=\begin{pmatrix}
K_M & \textbf{0}_{n_M\times n_E} \\
\textbf{0}_{n_E\times n_M} & I_{n_E\times n_E}
\end{pmatrix}
\end{align}
Here $I_{n_E\times n_E}$ is the $n_E\times n_E$ identity matrix,
and $K_M$ is a $n_M\times n_M$ orthogonal matrix. With this form
of matrix $K$ it is made clear that the $\bar{S}_r$'s actually
don't get transformed in this step, however the
$-\frac{1}{2}\xi^p\tilde{G}_p^2$ term is diagonalized by $K_M$.

It remains to find the $n_M\times n_M$ orthogonal matrix $K_M$. We note that
$\mL'$ written in the form of Eq.~\eqref{eq:lp}
is highly suggestive, because it has already completed the square. Therefore it
seems natural to guess that the transformation we need is simply
\begin{equation}
\tilde{S}_p=\alpha_p \tilde{G}_p,\quad p=1,2,...,n_M\,\text{(no summation over $p$)}
\label{eq:st}
\end{equation}
Here the $\alpha_p$'s are constants chosen to make the transformed fields
canonically normalized. Because the $\tilde{G}_p$'s can be
expressed as linear combinations of $\bar{S}_i,i=1,2,...,n_M$,
Eq.~\eqref{eq:st} effectively leads to a transformation from
$\bar{S}_i,i=1,2,...,n_M$ to $\tilde{S}_i,i=1,2,...,n_M$, from
which the matrix $K_M$ can be inferred.

There is one remaining potential loophole that
we need to deal with. It is necessary to ensure that the matrix $K_M$
inferred from Eq.~\eqref{eq:st} is indeed an orthogonal matrix,
otherwise we will not be able to keep the scalar kinetic terms in
their diagonalized and canonically normalized form.

To help determine whether the matrix $K_M$ inferred from
Eq.~\eqref{eq:st} is orthogonal we denote the real vector space spanned
by $G_i,i=1,2,...,n_S$ as $\mathbb{L}$ and introduce an
inner product in $\mathbb{L}$, defined by
\begin{equation}
\langle S_i|S_j\rangle\equiv\delta_{ij},i,j=1,2,...,n_S
\end{equation}
This means the $S_i$'s constitute an orthonormal basis in $\mathbb{L}$.
The inner product of any two elements in $\mathbb{L}$ can then
be calculated by virtue of the linearity property of the
inner product. It is obvious that the $\bar{S}_i$'s also
form an orthonormal basis in $\mathbb{L}$. Based on simple
algebraic knowledge the problem of judging whether $K_M$
is orthogonal reduces to judging whether $\tilde{S}_p,p=1,2,...,n_M$
form an orthonormal basis in the subspace spanned by themselves.

As long as all the $\tilde{G}_p$'s have positive norm,
we may always adjust the $\alpha_p$'s so that
\begin{equation}
\langle\tilde{S}_p|\tilde{S}_p\rangle=1,\quad\forall p=1,2,...,n_M
\end{equation}
Therefore the question becomes whether
$\langle\tilde{S}_p|\tilde{S}_q\rangle=0$ holds when
$p,q=1,2,...,n_M$ and $p\neq q$.
According to Eq.~\eqref{eq:st} we only need to check
whether $\langle\tilde{G}_p|\tilde{G}_q\rangle=0$
holds when $p,q=1,2,...,n_M$ and $p\neq q$.

Fortunately, when the scalar
fields are canonically normalized in their kinetic part,
the vector-scalar two-point transitions in a gauge theory
has the form~\cite{Weinberg:1996kr}
\begin{equation}
i\sum_{nm\alpha}\partial_\mu\phi'_n t_{nm}^\alpha A_{\alpha}^{\mu}v_m
\label{eq:cvs}
\end{equation}
Here $\phi'_n$ is the shifted scalar field with zero vacuum
expectation value, $v_m$ is the vacuum expectation value of
the original scalar fields. $t^\alpha$ denotes the generator
matrix with $\alpha$ being the adjoint index and
$A_\alpha^\mu$ is the corresponding gauge field. On the
other hand, the elements of the gauge boson mass matrix
are~\cite{Weinberg:1996kr}
\begin{equation}
\mu_{\alpha\beta}^2=-\sum_{nml}t_{nm}^\alpha t_{nl}^\beta v_m v_l
\label{eq:cvv}
\end{equation}
Compare Eq.~\eqref{eq:cvs} and Eq.~\eqref{eq:cvv} it is easy
to find for our case the useful property
\begin{equation}
\langle\tilde{G}_p|\tilde{G}_q\rangle=(\bM_V^2)_{pq},\forall p,q=1,2,...,n_M
\label{eq:gip}
\end{equation}
A nonlinearly-realized scalar sector does not introduce additional
difficulty in arriving at Eq.~\eqref{eq:gip}, because compared to
the linearly-realized case, the relevant differences begin from
quadratic terms in the field expansion and do not affect
Eq.~\eqref{eq:cvs} and Eq.~\eqref{eq:cvv}.

Eq.~\eqref{eq:gip} suggests that if the gauge bosons are already
in their mass eigenstates, then the related Goldstone boson vectors
must be orthogonal to each other, which is exactly what we desire.
Physically this implies that massive gauge bosons eat their
corresponding Goldstone bosons along the directions dictated by
their mass eigenstates. Therefore it would be desirable we rotate the gauge boson fields
to their mass eigenstates before adding the gauge-fixing terms
Eq.~\eqref{eq:gf}. This offers great convenience for the diagonalization
of scalar mass matrix afterwards.

On the other hand, if the gauge-fixing terms in Eq.~\eqref{eq:gf} are
added when $Z_p^\mu$'s are not mass eigenstate fields, although this
way of gauge-fixing is also legitimate, it would cause further
inconveniences. First, after rotation to gauge boson mass eigenstates,
the term $-\frac{1}{2\xi^p}(\partial_\mu Z_p^\mu)^2$ will induce
kinetic mixing between gauge bosons in a general $R_\xi$ gauge,
spoiling the diagonalization of gauge boson kinetic terms. Secondly,
from Eq.~\eqref{eq:gip} it is obvious that now the $\tilde{G}_p$'s
are not orthogonal to each other. Therefore the diagonalization
of scalar mass terms would not be straightforward. Due to the above
considerations in the following we adopt the procedure in which
gauge-fixing terms Eq.~\eqref{eq:gf} are added after rotating
gauge boson fields to their mass eigenstates.

Suppose the gauge boson mass matrix $\bM_V^2$ can be diagonalized
as follows
\begin{equation}
R\bM_V^2 R^{-1}=\bM_{DV}^2\equiv\text{diag}\{\mu_1^2,\mu_2^2,...,\mu_{n_M}^2\}
\end{equation}
Here $R$ is a $n_M\times n_M$ orthogonal matrix, and $\mu_1^2,\mu_2^2,...,\mu_{n_M}^2$
are positive. Let us define
\begin{widetext}
\begin{equation}
G_p^m\equiv\frac{R_{pq}}{\mu_p}\tilde{G}_q=\frac{(RF)_{pi}}{\mu_p}G_i,\quad p=1,2,...,n_M\,\text{(no summation over $p$)}
\end{equation}
\end{widetext}
(superscript $m$ denotes canonically-normalized mass eigenstates).
Now we can check with the help of Eq.~\eqref{eq:gip}(no summation over
$p,q$)
\begin{align}
\langle G_p^m|G_q^m\rangle=\frac{1}{\mu_p\mu_q}(R\bM_V^2 R^T)_{pq}=\delta_{pq},\forall p,q=1,2,...,n_M
\end{align}

We could further extend the definition of $G_p^m$ to the states $\bar{S}_r,r=n_M+1,...,n_S$ which
we have already obtained. According to our diagonalization of physical
scalar mass term, $\bar{S}_r$ can be expressed as
\begin{equation}
\bar{S}_r=(PU)_{ri}G_i,r=n_M+1,...,n_S
\end{equation}
where the matrix $U$ and $P$ are
introduced in Eq.~\eqref{eq:sug} and
Eq.~\eqref{eq:pt}, respectively. Finally we can
express $G_i^m$ as follows
\begin{equation}
G_i^m=Q_{ij}G_j,i=1,2,...,n_S
\label{eq:Q1}
\end{equation}
where the $n_S\times n_S$ matrix $Q$ is defined by (no summation over $i$)
\begin{align}
Q_{ij}=\begin{cases}
\frac{(RF)_{ij}}{\mu_i}, & i=1,2,...,n_M, \\
(PU)_{ij}, & i=n_M+1,...,n_S.
\end{cases}
\label{eq:Q2}
\end{align}
With the transformation matrix $R$ and $Q$ at our hand it will then be straightforward
to derive any three-point or four-point interaction that we are interested in.

\section{The case of SLH}
\label{sec:slh}

\subsection{Preparation for the calculation}

The SLH model was proposed as a simple solution to the Higgs mass naturalness problem,
making use of the collective symmetry breaking mechanism~\cite{ArkaniHamed:2002qy}. Its electroweak gauge group is
enlarged to $SU(3)_L\times U(1)_X$, and two scalar triplets are introduced to realize
the global symmetry breaking pattern
\begin{widetext}
\begin{equation}
[SU(3)_1\times U(1)_1]\times[SU(3)_2\times U(1)_2]
\to[SU(2)_1\times U(1)_1]\times[SU(2)_2\times U(1)_2]
\end{equation}
\end{widetext}
The scalar sector of the SLH model is usually written in a nonlinearly-realized
form. In this paper we follow the convention of ~\cite{delAguila:2011wk} and
parameterize the two scalar triplets as follows
\begin{align}
\Phi_1=\exp\left(\frac{i\Theta'}{f}\right)
\exp\left(\frac{it_\beta\Theta}{f}\right)
\begin{pmatrix}
0 \\ 0 \\ fc_\beta
\end{pmatrix} \label{eq:phi1} \\
\Phi_2=\exp\left(\frac{i\Theta'}{f}\right)
\exp\left(-\frac{i\Theta}{ft_\beta}\right)
\begin{pmatrix}
0 \\ 0 \\ fs_\beta
\end{pmatrix} \label{eq:phi2}
\end{align}
Here we introduced the shorthand notation
$s_\beta\equiv\sin\beta,c_\beta\equiv\cos\beta,t_\beta\equiv\tan\beta$.
$f$ is the Goldstone decay constant which is supposed to be
at least a few$\TeV$. $\Theta$ and $\Theta'$ are $3\times 3$ matrix fields, defined by
\begin{align}
\Theta=\frac{\eta}{\sqrt{2}}+
\begin{pmatrix}
\textbf{0}_{2\times 2} & h \\
h^\dagger & 0
\end{pmatrix},\quad
\Theta'=\frac{\zeta}{\sqrt{2}}+
\begin{pmatrix}
\textbf{0}_{2\times 2} & k \\
k^\dagger & 0
\end{pmatrix}
\end{align}
where $h$ and $k$ are parameterized as ($v\approx 246\GeV$ denotes
the vacuum expectation value of the Higgs doublet)
\begin{align}
h & =\begin{pmatrix} h^0 \\ h^- \end{pmatrix},\quad
h^0=\frac{1}{\sqrt{2}}(v+H-i\chi) \\
k & =\begin{pmatrix} k^0 \\ k^- \end{pmatrix},\quad
k^0=\frac{1}{\sqrt{2}}(\sigma-i\omega)
\end{align}
The covariant derivative in the electroweak sector
can be written as
\begin{equation}
D_\mu=\partial_\mu-igA_\mu^a T^a+ig_xQ_xB_\mu^x,\quad
g_x=\frac{gt_W}{\sqrt{1-t_W^2/3}}
\end{equation}
Here $t_W\equiv\tan\theta_W$.$A_\mu^a$ and $B_\mu^x$
denote the $SU(3)_L$ and $U(1)_X$ gauge fields,
respectively. The $SU(3)_C\times SU(3)_L\times U(1)_X$
gauge quantum number of $\Phi_1,\Phi_2$ is
$(\textbf{1},\textbf{3})_{-\frac{1}{3}}$, therefore
for $\Phi_1,\Phi_2$, $Q_x=-\frac{1}{3}$, and $A_\mu^a T^a$
can be written as
\begin{widetext}
\begin{align}
A_\mu^a T^a=\frac{A_\mu^3}{2}
\begin{pmatrix}
1 & 0 & 0 \\
0 & -1 & 0 \\
0 & 0 & 0
\end{pmatrix}
+\frac{A_\mu^8}{2\sqrt{3}}
\begin{pmatrix}
1 & 0 & 0 \\
0 & 1 & 0 \\
0 & 0 & -2
\end{pmatrix}
+\frac{1}{\sqrt{2}}
\begin{pmatrix}
0 & W_\mu^+ & Y_\mu^0 \\
W_\mu^- & 0 & X_\mu^- \\
Y_\mu^{0\dagger} & X_\mu^+ & 0
\end{pmatrix}
\end{align}
\end{widetext}
The gauge kinetic terms for $\Phi_1,\Phi_2$ are
\begin{equation}
\mL_{gk}=(D_\mu\Phi_1)^\dagger(D^\mu\Phi_1)+
(D_\mu\Phi_2)^\dagger(D^\mu\Phi_2)
\end{equation}
The first order (in $\frac{v}{f}$) gauge boson mixing
for $A^3,A^8,B_x$ takes the form
\begin{align}
\begin{pmatrix}
A^3 \\ A^8 \\ B_x
\end{pmatrix}
=
\begin{pmatrix}
0 & c_W & -s_W \\
\sqrt{1-\frac{t_W^2}{3}} & \frac{s_W t_W}{\sqrt{3}} & \frac{s_W}{\sqrt{3}} \\
-\frac{t_W}{\sqrt{3}} & s_W\sqrt{1-\frac{t_W^2}{3}} & c_W\sqrt{1-\frac{t_W^2}{3}}
\end{pmatrix}
\begin{pmatrix}
Z' \\ Z \\ A
\end{pmatrix}
\end{align}
We note that $Z',Z$ are not the ultimate mass eigenstate fields. For future
convenience we split the $Y^0$ field into real and imaginary parts
\begin{equation}
Y_\mu^0\equiv\frac{1}{\sqrt{2}}(Y_{R\mu}+iY_{I\mu}),\quad
Y_\mu^{0\dagger}\equiv\frac{1}{\sqrt{2}}(Y_{R\mu}-iY_{I\mu})
\end{equation}
In this paper we intend to focus on the neutral sector, in which there
are six scalar degrees of freedom: $\eta,\zeta,H,\chi,\sigma,\omega$.
Four degrees of freedom will be eaten to give mass to massive neutral
gauge bosons and are unphysical. The remaining two are physical and
need to play the role of the observed Higgs-like boson and the
pseudo-axion which has been discussed a lot in the literature. The pseudo-axion
actually corresponds to the Goldstone boson of a spontaneously broken
global $U(1)$ symmetry in the SLH. To give it a mass, the so-called
`$\mu$ term' needs to be introduced
\begin{equation}
\mL_\mu=\mu^2(\Phi_1^\dagger\Phi_2+\text{h.c.})
\end{equation}
The observed Higgs-like boson will acquire its mass from the Coleman-Weinberg
potential (however the $\mu$ term will also contribute to
its potential). Because $\mL_{gk},\mL_\mu$ and the Coleman-Weinberg potential
conserve CP, it will be convenient to group the neutral bosons into the CP-even and CP-odd
sectors: $H,\sigma,Y_R$ belong to the CP-even sector, while
$\eta,\zeta,\chi,\omega,Z',Z,Y_I,A$ belong to the CP-odd sector. There are no two-point
transitions between these two sectors.

Some comments concerning the parametrization of $\Phi_1,\Phi_2$ in
Eq.~\eqref{eq:phi1} and Eq.~\eqref{eq:phi2} are in order. Firstly, we
have chosen to retain the heavy sector fields in $\Theta'$, rather than
omitting them from the beginning. Apparently the omission of $\Theta'$
can be justified by doing a $SU(3)_L$ gauge transformation. This justification
is valid, and in the more precise language of Faddeev-Popov gauge-fixing,
the omission of $\Theta'$ actually corresponds to a certain choice of the
gauge-fixing function. However, this omission could lead to future
inconvenience, since as we will show, $\mL_{gk}$ contains two-point transitions
between heavy sector gauge bosons and the pseudo-axion. $\Theta'$ can
be rotated away by a gauge transformation but heavy sector gauge bosons
cannot. This means that when doing perturbation theory we need to always
carry those two-point vector-scalar transitions, which are quite inconvenient.
Nevertheless, the omission of $\Theta'$ and heavy sector gauge bosons can
indeed be convenient if we only need to obtain the $\ord(\frac{v}{f})$ coefficient
of the mass eigenstate $ZH\eta$ vertex, since the effect of those omitted
two-point vector-scalar transitions will be suppressed due to the heavy
gauge boson masses. Secondly, we have chosen to parameterize $\Phi_1,\Phi_2$
with two exponentials for each, rather than use a single exponential like
\begin{align}
\Phi_{1,SE}=\exp\left[\frac{i}{f}(\Theta'+t_\beta\Theta)\right]
\begin{pmatrix}
0 \\ 0 \\ fc_\beta
\end{pmatrix}
\end{align}
Also, in Eq.~\eqref{eq:phi1} and Eq.~\eqref{eq:phi2} the exponential
of $\Theta'$ has been put to the left of the exponential of $\Theta$.
For noncommutative matrices the single exponential parametrization
is not mathematically equivalent to the double exponential
parametrization. Moreover, the double exponential parametrization
will depend on the order of the two exponentials. However, these
parametrizations are related to each other by field redefinition
and should thus be physically equivalent. Which one to use is a
matter of convenience. We choose the double exponential
parametrization in Eq.~\eqref{eq:phi1} and Eq.~\eqref{eq:phi2}
because it does not introduce mass mixing between heavy and light
sector scalars in $\mathcal{L}_\mu$ and will thus facilitate the mass diagonalization.

The aim of this section is to derive the mass eigenstate $ZH\eta$
vertex in the SLH. With the current double exponential parametrization
it is possible to demonstrate that $H$ does not mix with $\sigma$,
and the scalar kinetic terms are already canonically-normalized
in the CP-even sector. Also,
the $\mu$ term gives $\eta$ a mass but does not introduce mass
mixing between $\eta$ and other fields. According to our argument
in the previous section this means that after all the diagonalization
procedure is completed, the whole effect on $\eta$ is supposed to
be a simple rescaling. This offers great convenience for the
derivation of the mass eigenstate $ZH\eta$ vertex. The needed rescaling factor can
be easily computed. Going back to the notation of Section~\ref{sec:sys},
the inner product between two Goldstone bosons $G_i$ and $G_j$ in
Eq.~\eqref{eq:quad} satisfies
\begin{widetext}
\begin{align}
\langle G_i|G_j\rangle=(U^{-1})_{ik}(U^{-1})_{jl}\langle S_k|S_l\rangle
=(U^{-1})_{ik}(U^{-1})_{jl}\delta_{kl}
=(U^{-1})_{ik}(U^{-1})_{jk}
=(V^{-1})_{ij}
\end{align}
\end{widetext}
We employ the convention that $\eta,\zeta,\chi,\omega$ correspond
to indices $1,2,3,4$ respectively, therefore
\begin{equation}
\langle\eta|\eta\rangle=(V^{-1})_{11}
\end{equation}
Consequently, the ultimate mass eigenstate field $\eta^m$ is
related to $\eta$ through
\begin{equation}
\eta=\sqrt{(V^{-1})_{11}}\eta^m
\label{eq:eta}
\end{equation}

To obtain the mass eigenstate $ZH\eta$ vertex, we also
need to know the component of $\eta^m$ in $\zeta,\chi,\omega$.
For the case of the SLH, let us denote the CP-odd sector elements of the matrix $F$
introduced in Eq.~\eqref{eq:quad} as
\begin{align}
F=\begin{pmatrix}
F_{Z\eta} & F_{Z\zeta} & F_{Z\chi} & F_{Z\omega} \\
F_{Z'\eta} & F_{Z'\zeta} & F_{Z'\chi} & F_{Z'\omega} \\
F_{Y\eta} & F_{Y\zeta} & F_{Y\chi} & F_{Y\omega}
\end{pmatrix}
\end{align}
(We assume for the CP-odd sector gauge boson mass matrix, the first, second and
third row/column correspond to $Z,Z',Y_I$, respectively.)
In the third row, $F_{Y\eta}$ denotes the coefficient of
the two-point transition $Y_I^\mu\partial_\mu\eta$ (similar
for $F_{Y\zeta},F_{Y\chi},F_{Y\omega}$). Due to CP-conservation
there is no two-point transition between $Y_R^\mu$ and the CP-odd
scalars, therefore no confusion would arise. The photon field $A^\mu$
does not have two-point transition with scalars. We would like
to denote the submatrix formed by the second, third and fourth
column of $F$ as $\tilde{F}$
\begin{align}
\tilde{F}\equiv\begin{pmatrix}
F_{Z\zeta} & F_{Z\chi} & F_{Z\omega} \\
F_{Z'\zeta} & F_{Z'\chi} & F_{Z'\omega} \\
F_{Y\zeta} & F_{Y\chi} & F_{Y\omega}
\end{pmatrix}
\end{align}
Now the application of Eq.~\eqref{eq:Q1} and Eq.~\eqref{eq:Q2}
to the CP-odd scalar sector of the SLH leads to
\begin{align}
\begin{pmatrix}
\zeta^m \\ \chi^m \\ \omega^m
\end{pmatrix}
=\mathbb{M}^{-1}_{DV}R\left[
\begin{pmatrix}
F_{Z\eta} \\ F_{Z'\eta} \\ F_{Y\eta}
\end{pmatrix}\eta
+\tilde{F}\begin{pmatrix}
\zeta \\ \chi \\ \omega
\end{pmatrix}
\right]
\label{eq:zetam}
\end{align}
As before the superscript $m$ denotes
canonically-normalized mass eigenstate fields. Inverting
Eq.~\eqref{eq:zetam} and using Eq.~\eqref{eq:eta}
will lead to
\begin{align}
\begin{pmatrix}
\zeta \\ \chi \\ \omega
\end{pmatrix}
=\tilde{F}^{-1}R^T\mathbb{M}_{DV}
\begin{pmatrix}
\zeta^m \\ \chi^m \\ \omega^m
\end{pmatrix}
-\sqrt{(V^{-1})_{11}}\tilde{F}^{-1}
\begin{pmatrix}
F_{Z\eta} \\ F_{Z'\eta} \\ F_{Y\eta}
\end{pmatrix}\eta^m
\label{eq:zeta}
\end{align}
We define the four-component column
vector
\begin{align}
\Upsilon\equiv
\begin{pmatrix}
\sqrt{(V^{-1})_{11}} \\
-\sqrt{(V^{-1})_{11}}\tilde{F}^{-1}
\begin{pmatrix}
F_{Z\eta} \\ F_{Z'\eta} \\ F_{Y\eta}
\end{pmatrix}
\end{pmatrix}
\label{eq:ups}
\end{align}
and denote the first row of $R$ as $\mathbb{R}_1$
\begin{equation}
\mathbb{R}_1=\begin{pmatrix} R_{11} & R_{12} & R_{13} \end{pmatrix}
\end{equation}
where $R_{ij}$ represents the $(i;j)$ element of $R$.
We will also need the coefficient matrices
\begin{widetext}
\begin{align}
\mathbb{C}^{dH}=\begin{pmatrix}
C^{dH}_{Z\eta} & C^{dH}_{Z\zeta} & C^{dH}_{Z\chi} & C^{dH}_{Z\omega} \\
C^{dH}_{Z'\eta} & C^{dH}_{Z'\zeta} & C^{dH}_{Z'\chi} & C^{dH}_{Z'\omega} \\
C^{dH}_{Y\eta} & C^{dH}_{Y\zeta} & C^{dH}_{Y\chi} & C^{dH}_{Y\omega}
\end{pmatrix},\quad
\mathbb{C}^{Hd}=\begin{pmatrix}
C^{Hd}_{Z\eta} & C^{Hd}_{Z\zeta} & C^{Hd}_{Z\chi} & C^{Hd}_{Z\omega} \\
C^{Hd}_{Z'\eta} & C^{Hd}_{Z'\zeta} & C^{Hd}_{Z'\chi} & C^{Hd}_{Z'\omega} \\
C^{Hd}_{Y\eta} & C^{Hd}_{Y\zeta} & C^{Hd}_{Y\chi} & C^{Hd}_{Y\omega}
\end{pmatrix}
\end{align}
\end{widetext}
Here $C^{dH}_{Z\eta}$ denotes the coefficient of $Z^\mu\eta\partial_\mu H$,
while $C^{Hd}_{Z\eta}$ denotes the coefficient of $Z^\mu H\partial_\mu\eta$,
and so on. If we have calculated the matrices
$\mathbb{C}^{dH},\mathbb{C}^{Hd}$ and the vectors $\Upsilon$ and $\mathbb{R}_1$,
then the coefficient of mass eigenstate antisymmetric $ZH\eta$ vertex
($Z^\mu(\eta\partial_\mu H-H\partial_\mu\eta)$ with all fields understood
to be mass eigenstate fields) can be obtained as
\begin{equation}
c^{as}_{ZH\eta}=\frac{\mathbb{R}_1 \mathbb{C}^{dH}\Upsilon
-\mathbb{R}_1 \mathbb{C}^{Hd}\Upsilon}{2}
\label{eq:cas}
\end{equation}
while the coefficient of mass eigenstate symmetric $ZH\eta$ vertex
($Z^\mu(\eta\partial_\mu H+H\partial_\mu\eta)$ with all fields understood
to be mass eigenstate fields) can be obtained as
\begin{equation}
c^{s}_{ZH\eta}=\frac{\mathbb{R}_1 \mathbb{C}^{dH}\Upsilon
+\mathbb{R}_1 \mathbb{C}^{Hd}\Upsilon}{2}
\label{eq:cs}
\end{equation}
Here we remark that we divide a general VSS vertex into its antisymmetric
and symmetric parts because they exhibit distinct features in physical
processes. For example, the symmetric VSS vertex does not contribute when
the involved vector boson is on shell. Therefore, only the antisymmetric
$ZH\eta$ vertex is expected to contribute at tree level to decay processes
$H\rightarrow Z\eta$ (or $\eta\rightarrow ZH$ if $\eta$ is heavy) where
$Z$ is supposed to be on shell.

\subsection{Results}

In principle the derivation of mass eigenstate $ZH\eta$ vertex
with no expansion on the $\frac{v}{f}$ can be carried out
manually\footnote{In practice, they can be more readily obtained
with the help of $\texttt{Mathematica}$.}. However, after obtaining $V,F$ and $\mathbb{M}_V^2$,
the calculation of $R$ and the inverse matrices can become
extremely cumbersome. Therefore we choose to compute the
mass eigenstate $ZH\eta$ vertex to $\ord ((\frac{v}{f})^3)$,
which makes the results easier to obtain and display. For brevity
we define $\xi\equiv\frac{v}{f}$ in the following.

Let us first find the scalar kinetic matrix $V$ and vector-scalar
transition matrix $F$ for the SLH. They are computed to be
\begin{widetext}
\begin{align}
V=\begin{pmatrix}
1 & 0 & \frac{\sqrt{2}}{t_{2\beta}}\xi-\frac{7c_{2\beta}+c_{6\beta}}{6\sqrt{2}s_{2\beta}^3}\xi^3 & -\sqrt{2}\xi+\frac{5+3c_{4\beta}}{3\sqrt{2}s_{2\beta}^2}\xi^3 \\
0 & 1 & -\frac{1}{\sqrt{2}}\xi+\frac{5+3c_{4\beta}}{12\sqrt{2}s_{2\beta}^2}\xi^3 & -\frac{2\sqrt{2}}{3t_{2\beta}}\xi^3 \\
\frac{\sqrt{2}}{t_{2\beta}}\xi-\frac{7c_{2\beta}+c_{6\beta}}{6\sqrt{2}s_{2\beta}^3}\xi^3 & -\frac{1}{\sqrt{2}}\xi+\frac{5+3c_{4\beta}}{12\sqrt{2}s_{2\beta}^2}\xi^3 & 1-\frac{5+3c_{4\beta}}{12s_{2\beta}^2}\xi^2 & \frac{2}{3t_{2\beta}}\xi^2 \\
-\sqrt{2}\xi+\frac{5+3c_{4\beta}}{3\sqrt{2}s_{2\beta}^2}\xi^3 & -\frac{2\sqrt{2}}{3t_{2\beta}}\xi^3 & \frac{2}{3t_{2\beta}}\xi^2 & 1
\end{pmatrix}+\ord(\xi^4)
\label{eq:vij}
\end{align}
\begin{align}
F=
gf\begin{pmatrix}
\frac{1}{\sqrt{2}c_W t_{2\beta}}\xi^2 & -\frac{1}{2\sqrt{2}c_W}\xi^2 & \frac{1}{2c_W}\xi-\frac{5+3c_{4\beta}}{24c_W s_{2\beta}^2}\xi^3 & \frac{1}{3c_W t_{2\beta}}\xi^3 \\
\frac{\rho}{t_{2\beta}}\xi^2 & \frac{\sqrt{2}}{\sqrt{3-t_W^2}}-\frac{1+2c_{2W}}{2\sqrt{2}c_W^2\sqrt{3-t_W^2}}\xi^2 & \kappa\xi-\frac{\kappa (5+3c_{4\beta})}{12s_{2\beta}^2}\xi^3 & -\frac{1}{3c_W^2\sqrt{3-t_W^2}t_{2\beta}}\xi^3 \\
-\xi+\frac{5+3c_{4\beta}}{6s_{2\beta}^2}\xi^3 & -\frac{2}{3t_{2\beta}}\xi^3 & \frac{\sqrt{2}}{3t_{2\beta}}\xi^2 & \frac{1}{\sqrt{2}}
\end{pmatrix}
+\ord(\xi^4)
\label{eq:fij}
\end{align}
\end{widetext}
where we defined
\begin{align}
\rho\equiv\sqrt{\frac{1+2c_{2W}}{1+c_{2W}}},\quad \kappa\equiv\frac{c_{2W}}{2c_W^2\sqrt{3-t_W^2}}
\end{align}
It is obvious from Eq.~\eqref{eq:vij} that the scalar kinetic terms in the original
$\eta,\zeta,\chi,\omega$ are not canonically normalized, and also obvious from
Eq.~\eqref{eq:fij} that there are general vector-scalar two-point transitions.
Especially, the two-point $Z\eta$ transition appears at $\ord(\xi^2)$, only
one order of $\xi$ relatively suppressed when compared to $Z\chi$ transition
\footnote{Although the two-point $Z\eta$ transition appears at $\ord(\xi^2)$,
the elimination of this part require an $\ord(\xi)$ field redefinition,
due to the fact that the relative suppression of $Z\eta$ transition to
$Z\chi$ transition is $\ord(\xi)$. The $ZH\chi$ coupling is $\ord(1)$.
Therefore, the removal of $Z\eta$ transition could lead to an $\ord(\xi)$
change in the derived $ZH\eta$ vertex.}. The appearance of these
non-canonically normalized kinetic terms and
`unexpected'\footnote{By `unexpected' we refer to the fact that $\eta$ is considered
physical, yet there exist two-point transitions such as $Z^\mu\partial_\mu\eta$
in $\mL_{gk}$.} vector-scalar transitions is the exact reason for introducing
the systematic procedure in Section~\ref{sec:sys}.

The $\Upsilon$ vector is computed to be
\begin{equation}
\Upsilon=\begin{pmatrix}
1+\frac{1}{s_{2\beta}^2}\xi^2+\ord(\xi^4) \\
-\frac{1}{t_{2\beta}}\xi^2+\ord(\xi^4) \\
-\frac{\sqrt{2}}{t_{2\beta}}\xi
-\frac{3-c_{4\beta}}{\sqrt{2}s_{2\beta}^2 t_{2\beta}}\xi^3+\ord(\xi^5) \\
\sqrt{2}\xi+\frac{3-c_{4\beta}}{3\sqrt{2}s_{2\beta}^2}\xi^3+\ord(\xi^5)
\end{pmatrix}
\label{eq:ups}
\end{equation}
A compact expression for $\Upsilon$ valid to all orders
in $\xi$ can also be obtained. It is
\begin{align}
\Upsilon=\begin{pmatrix}
c_{\gamma+\delta}^{-1} \\
\\
-c_{\gamma+\delta}^{-1}(s_\delta^2 t_\beta-s_\gamma^2 t_\beta^{-1}) \\
\\
\frac{v}{\sqrt{2}f}c_{\gamma+\delta}^{-1}(c_{2\delta}t_\beta-c_{2\gamma}t_\beta^{-1}) \\
\\
\frac{1}{2}c_{\gamma+\delta}^{-1}(s_{2\delta}t_\beta+s_{2\gamma}t_\beta^{-1})
\end{pmatrix}
\end{align}
where
\begin{align}
\gamma\equiv\frac{vt_\beta}{\sqrt{2}f},\quad \delta\equiv\frac{v}{\sqrt{2}ft_\beta}
\end{align}

Expanding the above expression to $\mathcal{O}(\xi^3)$, Eq.~\eqref{eq:ups}
can be recovered. The above expression for the $\Upsilon$ vector
is very useful in derivation of exact results of tree level
vertices involving the $\eta$ particle. The $\mathbb{C}^{dH}$ matrix is computed to be
\begin{widetext}
\begin{equation}
\mathbb{C}^{dH}=\begin{pmatrix}
0 & 0 & -\frac{g}{2c_W}+\frac{g(5+3c_{4\beta})}{24c_W s_{2\beta}^2}\xi^2+\ord(\xi^4) & 0 \\
0 & 0 & -\frac{g(1-t_W^2)}{2\sqrt{3-t_W^2}}+\frac{g\kappa (5+3c_{4\beta})}{12s_{2\beta}^2}\xi^2+\ord(\xi^4) & 0 \\
0 & 0 & -\frac{\sqrt{2}g}{3t_{2\beta}}\xi+\frac{g(7c_{2\beta}+c_{6\beta})}{30\sqrt{2}s_{2\beta}^3}\xi^3+\ord(\xi^5) & 0
\end{pmatrix}
\end{equation}
The $\mathbb{C}^{Hd}$ matrix is computed to be
\begin{align}
\mathbb{C}^{Hd}=
\begin{pmatrix}
\frac{\sqrt{2}g}{c_W t_{2\beta}}\xi-\frac{g(7c_{2\beta}+c_{6\beta})}{3\sqrt{2}c_W s_{2\beta}^3}\xi^3 &
-\frac{g}{\sqrt{2}c_W}\xi+\frac{g(5+3c_{4\beta})}{6\sqrt{2}c_W s_{2\beta}^2}\xi^3 &
\frac{g}{2c_W}-\frac{g(5+3c_{4\beta})}{8c_W s_{2\beta}^2}\xi^2 &
\frac{g}{c_W t_{2\beta}}\xi^2 \\
\frac{2g\rho}{t_{2\beta}}\xi-\frac{g\rho(7c_{2\beta}+c_{6\beta})}{3s_{2\beta}^3}\xi^3
& -g\rho\xi+\frac{g\rho(5+3c_{4\beta})}{6s_{2\beta}^2}\xi^3
& g\kappa-\frac{g\kappa(5+3c_{4\beta})}{4s_{2\beta}^2}\xi^2
& -\frac{g}{c_W^2\sqrt{3-t_W^2}t_{2\beta}}\xi^2 \\
-g+\frac{g(5+3c_{4\beta})}{2s_{2\beta}^2}\xi^2
& -\frac{2g}{t_{2\beta}}\xi^2
& \frac{2\sqrt{2}g}{3t_{2\beta}}\xi-\frac{\sqrt{2}g(7c_{2\beta}+c_{6\beta})}{15s_{2\beta}^3}\xi^3
& 0
\end{pmatrix}+\ord(\xi^4)
\end{align}
The matrix $R$ can be computed as
\begin{align}
R=\begin{pmatrix}
1+\ord(\xi^4) & -\frac{c_{2W}(1+2c_{2W})}{8c_W^5\sqrt{3-t_W^2}}\xi^2+\ord(\xi^4) & -\frac{\sqrt{2}}{3c_W t_{2\beta}}\xi^3+\ord(\xi^5) \\
\frac{c_{2W}(1+2c_{2W})}{8c_W^5\sqrt{3-t_W^2}}\xi^2+\ord(\xi^4) & 1+\ord(\xi^4) & -\frac{\sqrt{2}(1+2c_{2W})}{3c_W^2\sqrt{3-t_W^2}t_{2\beta}}\xi^3+\ord(\xi^5) \\
\frac{\sqrt{2}}{3c_W t_{2\beta}}\xi^3+\ord(\xi^5) & \frac{\sqrt{2}(1+2c_{2W})}{3c_W^2\sqrt{3-t_W^2}t_{2\beta}}\xi^3+\ord(\xi^5) & 1+\ord(\xi^6)
\end{pmatrix}
\end{align}
With this precision it is feasible to obtain $c^{as}_{ZH\eta}$ and $c^{s}_{ZH\eta}$ via
Eq.~\eqref{eq:cas} and Eq.~\eqref{eq:cs} to $\ord(\xi^3)$, the results of which are
\begin{align}
c^{as}_{ZH\eta}=-\frac{g}{4\sqrt{2}c_W^3 t_{2\beta}}\xi^3+\ord(\xi^5)
\label{eq:casv}
\end{align}
\begin{align}
c^{s}_{ZH\eta}=\frac{g}{\sqrt{2}c_W t_{2\beta}}\xi
+\frac{g}{24\sqrt{2}c_W s_{2\beta}}\left[\frac{8}{s_{2\beta}t_{2\beta}}+3c_{2\beta}\left(8+\frac{6}{c_W^2}-\frac{1}{c_W^4}\right)\right]\xi^3
+\ord(\xi^5)
\label{eq:csv}
\end{align}
\end{widetext}
Therefore we arrive at the conclusion that the symmetric $ZH\eta$ vertex appear at $\ord(\xi)$,
while the antisymmetric $ZH\eta$ vertex does not appear until $\ord(\xi^3)$. The coefficients
of these two vertices are presented in Eq.~\eqref{eq:csv} and Eq.~\eqref{eq:casv}, respectively.
We note that this conclusion differs from what has been derived and used in the
literature~\cite{Kilian:2004pp,Kilian:2006eh} for a long time. In the intermediate steps,
one important discrepancy between our results and Ref.~\cite{Kilian:2004pp} is that
in a footnote Ref.~\cite{Kilian:2004pp} claims that choosing the $\eta$ generator to be
the identity matrix would remove the kinetic mixing between $\eta$ and unphysical Goldstone
bosons, while in our derivation Eq.~\eqref{eq:vij} shows there still exists $\ord(\xi)$
kinetic mixing of such kind, which we have checked by various means.
It is then not clear whether Ref.~\cite{Kilian:2004pp,Kilian:2006eh} have made appropriate
field redefinitions to diagonalize the SLH vector-scalar system.

\subsection{Effective Field Theory Analysis}

The fact that the mass eigenstate antisymmetric $ZH\eta$ vertex does not appear until $\ord(\xi^3)$
can be understood from an effective field theory (EFT) point of view. Let us focus on
the bosonic sector of the SLH, and integrate out heavy sector fields $X,Y,Z'$ and their
Goldstones. We are then interested in the EFT formed with the remaining fields, namely
the SM and $\eta$, which are classified according to gauge transformation properties.
Especially, $\eta$ is a singlet under the SM gauge symmetries. Let us suppose at this moment
we have not added the gauge-fixing terms yet. It is obvious that at dimension-four level
no gauge-invariant operator can deliver a $ZH\eta$ vertex. We are then forced to
consider higher-dimensional operators. At dimension-five level, let us consider
\begin{align}
\mathcal{O}_1=(\partial^\mu\eta)[ih^\dagger(D_\mu-\overleftarrow{D_\mu})h]
\end{align}
where $h^\dagger\overleftarrow{D_\mu}h\equiv(D_\mu h)^\dagger h$ and $D_\mu$
denotes the SM covariant derivative for the Higgs doublet. We may denote its coefficient
as $\frac{c_1}{f}$, in which $c_1$ is a dimensionless constant.
Then we could find in the Lagrangian the following terms
\begin{widetext}
\begin{align}
\mL & \supset(D_\mu h)^\dagger(D^\mu h)+\frac{1}{2}(\partial_\mu\eta)^2+\frac{c_1}{f}\mathcal{O}_1 \nonumber\\
    & \supset\frac{1}{2}(\partial_\mu H)^2+\frac{1}{2}(\partial_\mu\chi)^2+\frac{1}{2}(\partial_\mu\eta)^2
    +\frac{v}{f}c_1(\partial^\mu\eta)(\partial_\mu\chi) \nonumber\\
    & -m_Z Z_\mu\partial^\mu\left(\chi+\frac{v}{f}c_1\eta\right)
    +\frac{m_Z}{v}Z_\mu(\chi\partial^\mu H-H\partial^\mu\chi)
    -\frac{2m_Z}{f}c_1 HZ_\mu\partial^\mu\eta
\end{align}
\end{widetext}
The appearance of scalar kinetic mixing $(\partial^\mu\eta)(\partial_\mu\chi)$
and vector-scalar two-point transition $Z_\mu\partial^\mu\eta$ signal the need
for a further field redefinition in the scalar sector. Up to $\ord(\xi)$, the transformation
is easily found:
\begin{align}
\tilde{\chi} & =\chi+\frac{v}{f}c_1\eta, \\
\tilde{\eta} & =\eta.
\label{eq:fr1}
\end{align}
The Lagrangian can be written with the transformed fields
\begin{align}
&\mL\supset\frac{1}{2}(\partial_\mu H)^2+\frac{1}{2}(\partial_\mu\tilde{\chi})^2
+\frac{1}{2}(\partial_\mu\tilde{\eta})^2-m_Z Z_\mu\partial^\mu\tilde{\chi}
 \nonumber\\&+\frac{m_Z}{v}Z_\mu(\tilde{\chi}\partial^\mu H-H\partial^\mu\tilde{\chi})
-c_1\frac{m_Z}{f}Z_\mu(\tilde{\eta}\partial^\mu H+H\partial^\mu\tilde{\eta})
\end{align}
The two-point vector-scalar transition $-m_Z Z_\mu\partial^\mu\tilde{\chi}$
can be eliminated by an appropriate $R_\xi$ gauge-fixing term. From the above
expression we see that at $\ord(\xi)$, only symmetric mass eigenstate $ZH\eta$
vertex could survive while the antisymmetric counterpart is removed after the
transition to mass eigenstate. This is similar to the situation considered
in Ref.~\cite{Bauer:2016zfj} which also concluded for the case of the SM plus
a singlet scalar $S$ that the dimension-five operator cannot give rise to
tree-level $S\rightarrow ZH$ decay.

At dimension-six level, let us consider the operator
\begin{align}
\mathcal{O}_2=(h^\dagger D^\mu h)(h^\dagger D_\mu h)
\end{align}
This operator should have a coefficient of $\ord\left(\frac{1}{f^2}\right)$.
Apparently it does not contain $\eta$. However, if $\mathcal{O}_1$ is
also present, then a field redefinition like Eq.~\eqref{eq:fr1} needs
to be performed, after which $\mathcal{O}_2$ could lead to a
mass eigenstate antisymmetric $ZH\eta$ vertex. Since the field redefinition
implies an $\ord(\xi)$ $\eta$ component in $\chi$, the resultant
mass eigenstate antisymmetric $ZH\eta$ vertex should appear at $\ord(\xi^3)$.

We may also consider operators with even higher dimension, but of course they
cannot lead to $\ord(\xi)$ or $\ord(\xi^2)$ mass eigenstate antisymmetric
$ZH\eta$ vertex.

Other bosonic operators (containing $Z$) at dimension-five or six level can be considered, for example
\begin{align}
\mathcal{O}_3 & =\eta(D_\mu h)^\dagger(D^\mu h) \\
\mathcal{O}_4 & =\partial^\mu(h^\dagger h)[ih^\dagger(D_\mu-\overleftarrow{D_\mu})h]
\end{align}
However, these operators do not have the correct CP property. Furthermore,
in our parametrization $\eta$ has a shift symmetry $\eta\rightarrow\eta+c$
where $c$ is a constant, which also forbids the appearance of $\mathcal{O}_3$.

Therefore from an EFT analysis, we also arrive at the conclusion that
in the SLH, mass eigenstate antisymmetric $ZH\eta$ vertex cannot appear
until $\ord(\xi^3)$ while symmetric $ZH\eta$ vertex can appear at $\ord(\xi)$
~\footnote{According to Ref.~\cite{Goh:2006wj}, a similar situation occurs
for the $ZH\phi_0$ vertex in the left-right twin Higgs model, where $\phi_0$
denotes a neutral pseudoscalar. This is consistent with our EFT analysis here,
since $\phi_0$ does not mix with other physical fields due to an imposed
discrete symmetry.}, consistent with our explicit calculation in the previous subsection. It is
important to note that all of the EFT derivation is based on the field content
SM$+\eta$ ($\eta$ is a CP-odd singlet~\footnote{Ref.~\cite{DeCurtis:2016tsm}
studied the composite two-Higgs-doublet model which contains $\ord(1)$ antisymmetric
$ZHA$ vertex since the pseudoscalar $A$ is not a singlet.}), with no additional particles leading to further mass mixings, which
could alter the conclusion.

\section{Discussion and conclusion}
\label{sec:conc}

In this paper we revisited the issue of deriving the mass eigenstate $ZH\eta$ vertex
in the SLH. We found that the scalar kinetic terms are not canonically normalized
in the usual parametrization and there are `unexpected'
vector-scalar two-point transitions that need to be taken care of. We formulated
the problem in a generic setting as the diagonalization of a vector-scalar system
in gauge field theories. Especially we proved that the scalar mass terms coming
from the $R_\xi$ gauge-fixing procedure will be automatically orthogonal to
each other if the corresponding gauge fields are rotated to their mass eigenstate
prior to gauge-fixing~\footnote{We refer the reader
to Ref.~\cite{Hubisz:2005tx} for another example in the Littlest Higgs
with T-parity.}. This fact greatly simplifies the diagonalization procedure.

For the SLH model, we found that the double exponential parametrization of
scalar triplets, as shown in Eq.~\eqref{eq:phi1} and Eq.~\eqref{eq:phi2}
is convenient for the derivation of $ZH\eta$ vertex, since in this parametrization
the $\eta$ field is only subject to a simple rescaling in the diagonalization
procedure, with which we could display in a simple form the $\eta^m$ component
contained in the original $\eta,\zeta,\chi,\omega$ fields we started with,
as shown in Eq.~\eqref{eq:ups}.

In principle the derivation of mass eigenstate $ZH\eta$ vertex could be worked out
to all order in $\xi\equiv\frac{v}{f}$, however the intermediate results are
too lengthy and we find it convenient to display the derivation and results
to $\ord(\xi^3)$. The final results of antisymmetric and symmetric $ZH\eta$
vertices are shown in Eq.~\eqref{eq:casv} and Eq.~\eqref{eq:csv}. Contrary to
what has existed in the literature~\cite{Kilian:2004pp,Kilian:2006eh} (which
claims an $\mathcal{O}(\xi)$ antisymmetric $ZH\eta$ vertex) for a long time
, we found that the coefficient
of the antisymmetric $ZH\eta$ vertex $c^{as}_{ZH\eta}$ does not show up
until $\ord(\xi^3)$. This result is also understood from an EFT point of view.
Based on these results we expect that the exotic Higgs
decay $H\rightarrow Z\eta$ (or $\eta\rightarrow ZH$ if $\eta$ is heavy)
and the associated production of $h$ and $\eta$ at hadron or lepton colliders will be
much more difficult to observe due to the $\mathcal{O}(\xi^3)$ suppression
in the antisymmetric $ZH\eta$ vertex. On the
other hand, the symmetric $ZH\eta$ vertex already appears at $\ord(\xi)$,
however the investigation of its effect involves some
subtleties, which will be treated in a follow-up paper.

The procedure elucidated in this paper can be applied to other models containing
a gauged nonlinearly-realized scalar sector as well. From the experience with
the SLH we find it important to examine the quadratic part of the Lagrangian
in these models, which could contain non-canonically normalized scalar kinetic
terms and `unexpected' vector-scalar two-point transitions. Moreover, finding
a convenient parametrization for the exponentials in these models could be
very helpful in the diagonalization procedure. We expect to investigate these
issues and their phenomenological implications in the future.

\subsection*{Acknowledgements}

We thank Kingman Cheung for helpful discussion. We also thank the
referee who drew our attention to an EFT viewpoint. This work was
supported in part by the Natural Science Foundation of China (Grants
No. 11135003, No. 11375014 and No. 11635001), and the China
Postdoctoral Science Foundation (Grant No. 2017M610992).

\bibliography{slhvssPRD}
\bibliographystyle{h-physrev}

\end{document}